\documentclass{aa}
\usepackage{graphicx}
\begin{document}
   \title{High-resolution spectroscopy of the low-mass 
             X-ray binary \object{EXO~0748$-$67}}
   \author{J. Cottam \inst{1}, S. M. Kahn \inst{1}, 
          A. C. Brinkman \inst{2}, J. W. den Herder\inst{2},  
          \and
          C. Erd\inst{3} 
          }

   \authorrunning {Cottam et al.}
   \titlerunning {High-Resolution Spectroscopy of \object{EXO~0748$-$67}}

   \offprints{jcottam@astro.columbia.edu}

   \institute{Columbia Astrophysics Laboratory, 
              Columbia University,
              550 West 120th Street, New York, New York 10027, USA
              \and Space Research Organization of the Netherlands,
              Laboratory for Space Research, 
              Sorbonnelaan 2, 3584 CA Utrecht, The Netherlands 
	      \and Astrophysics Division, Space Science Department of ESA, 
              ESTEC, 2200 AG Noordwijk, the Netherlands 
             }

   \date{Received <date> / Accepted <date>}

\abstract{
We present initial results from observations of the low-mass X-ray binary \object{EXO~0748$-$67} with the Reflection Grating Spectrometer on board the {\it XMM-Newton} Observatory.  The spectra exhibit discrete structure due to absorption and emission from ionized neon, oxygen, and nitrogen.  We use the quantitative constraints imposed by the spectral features to develop an empirical model of the circumsource material.  This consists of a thickened accretion disk with emission and absorption in the plasma orbiting high above the binary plane.  This model presents challenges to current theories of accretion in X-ray binary systems.
} 

   \maketitle

\section{Introduction}

X-ray spectroscopic observations should provide a useful probe of the accretion processes in low-mass X-ray binaries (LMXB's).  The continuum emission from the bright central source is reprocessed as the photons are absorbed and re-emitted in the surrounding material.  The details of the resulting X-ray spectra are extremely sensitive to the physical conditions in the plasma and therefore provide an excellent way to constrain the circumstellar environment and the mass flow that fuels the X-ray emission.  Previous observatories have not had the spectral resolution necessary to resolve the discrete structure in the X-ray spectrum.  The Reflection Grating Spectrometer (RGS) on the recently launched {\it XMM-Newton} Observatory provides both a high resolving power and a large effective area making it uniquely sensitive to the diagnostic spectral features, particularly in the soft X-ray band that contains transitions from highly ionized charge states of the most abundant elements.     

In this letter we present the results of RGS observations of \object{EXO~0748$-$67}, a highly variable LMXB that was first discovered with {\it EXOSAT} (Parmar et al. \cite{Parmar86}).  Analysis of the {\it EXOSAT} light curves, which show deep eclipses with a 3.82 hour orbital period and a complex dipping structure, led to a derived inclination angle of $75\degr-82\degr$ (Parmar et al. \cite{Parmar86}).  The detection of type I X-ray bursts confirmed that the compact object is a neutron star (Gottwald et al. \cite{Gottwald}).  Observations with both {\it ASCA} (Thomas et al. \cite{Thomas}) and {\it ROSAT} (Schulz \cite{Schulz}) revealed structure in the soft X-ray spectrum, but the limited spectral resolution of these instruments made it impossible to distinguish whether this was due to absorption or emission features.  In the RGS observations that we present below, we find a spectrum that is rich in discrete structure including absorption and emission from ionized oxygen, nitrogen, and neon.  We use the available spectral diagnostics to construct a quantitative empirical model of the circumsource flow.
  
\section{Data Reduction}

The RGS covers the wavelength range of $5$ to $35\, {\rm \AA}$ with a resolution of $0.05\, {\rm \AA}$ (roughly constant across the band) and a peak effective area of $\sim 140\, {\rm cm^{2}}$ at $15\, {\rm \AA}$.  A complete description of the instrument can be found in den Herder et al. (\cite{denHerder}).  \object{EXO~0748$-$67} was observed repeatedly during both the comissioning and calibration phases of the mission.  We have chosen the two longest exposures for this analysis; the first was 49.3 ks on 2000 March 28, and the second was 44.8 ks on 2000 April 21.  The raw data were processed with the development version of the {\it XMM-Newton} Science Analysis Software (SAS).  The spectra were extracted by first applying a $30\arcsec$ wide spatial filter to the CCD image.  The surviving events were then plotted in dispersion channel vs. CCD pulse-height space and additional filters were applied to collect events from the different spectral orders.  Background spectra were extracted by applying the same spectral order filters to events from a spatial region that was offset from the source location on the CCD image. 

We assigned nominal wavelengths to each dispersion channel based on the geometry of the instrument and the pointing angles of the spacecraft.  The pointings were found to be offset by $46\arcsec$ for the first observation and $60\arcsec$ for the second using images from the European Photon Imaging Camera (EPIC).  We expect the wavelengths to be accurate to $\sim 0.010\,{\rm \AA}$.  The development version of the SAS did not yet have the capability to correct the events for aspect variations.  However, after excluding the initial part of each observation (8.6 ks of the first observation and 1.7 ks of the second), we find that the pointings are stable to less than $\sim 2\arcsec$.  Since this is much smaller than the resolution of the telescope, further aspect correction is unnecessary. 

We have determined the effective area for each exposure by applying the same extraction regions that were used with the data to the full response matrix, which includes all information on the efficiency of the instruments.  Based on our ground calibration we expect the uncertainty to be less than $\sim 10\%$ for wavelengths longer than $9\, {\rm \AA}$ and at most $\sim 20 \%$ at the shortest wavelengths.  We have calculated the flux for each observation using these effective area curves and the standard SAS exposure maps, which give exposure times for each wavelength bin.  We combined the first order spectra from both instruments for the two observations in order to maximize the statistical quality of our measurement.         

\section{Observational Features}

\subsection{Light Curve}

\begin{figure}
      \resizebox{8.5cm}{!}{\rotatebox{0}{\includegraphics{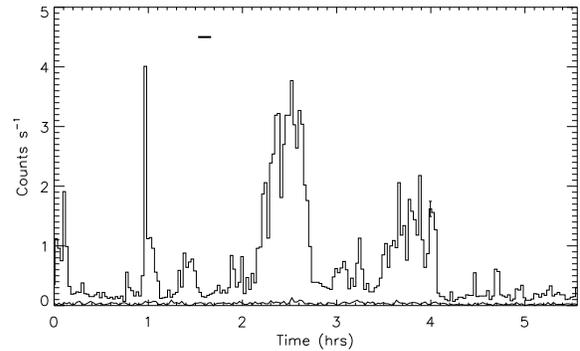}}}
      \caption{Partial first order light curve in $100\, {\rm s}$ bins.  The background level is indicated by the thick line.   A representative error bar (for statistical error only) is shown at $\sim 4 \, {\rm hr}$.  The horizontal bar marks the location of the hard X-ray eclipse.  The time units have an arbitrary origin.   
              }
         \label{FigLight}
   \end{figure}

The observed \object{EXO~0748$-$67} count rate varies by up to a factor of ten on timescales as short as a few hundred seconds.  This is illustrated in Fig. ~\ref{FigLight}, which shows a portion of the RGS light curve.  There are periods of stable low-level emission where the count rate is less than $\sim 0.5\, {\rm ct\, s^{-1}}$, periods where the count rate varies rapidly between $\sim 0.5$ and $3.5\, {\rm ct\, s^{-1}}$, and several type I X-ray bursts with peak intensities from $2.5$ to $5.2\, {\rm ct\, s^{-1}}$.  We see no correlation between the variability and the orbital phase; similar to the soft X-ray light curve from {\it ASCA} (Church et al \cite{Church}), the RGS light curve does not show the quiescent level and dipping structure that is characteristic of the light curves from higher energy observations.  Most importantly, we see no eclipses during the times predicted for the hard X-ray eclipses.  
     
\subsection{Discrete Spectral Structure}

\begin{figure}
      \resizebox{8.5cm}{!}{\rotatebox{0}{\includegraphics{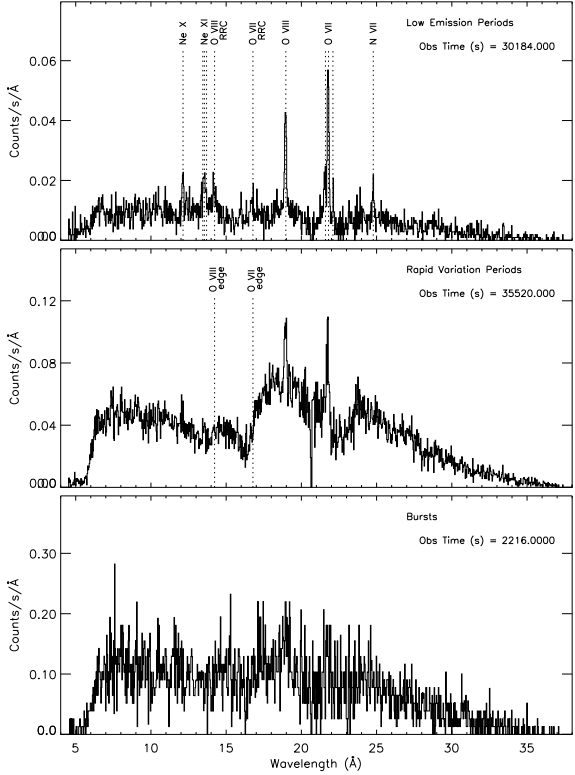}}}
      \caption[]{Spectra for the periods of low emission, active variation and burst as defined in Sect. 3.1.  The cumulative exposure time for each spectrum is indicated.   
              }
         \label{FigSpectra}
   \end{figure}

As shown in Fig. ~\ref{FigSpectra}, the \object{EXO~0748$-$67} spectra show significant discrete structure both in absorption and emission.  We see bright emission lines from \ion{O}{viii} Ly$\alpha$ and the \ion{O}{vii} He-like complex.  The \ion{Ne}{x} Ly$\alpha$, \ion{Ne}{ix} He-like complex and the \ion{N}{vii} Ly$\alpha$ emission lines are weaker, but clearly visible, particularly during the periods of low emission when the equivalent widths of the lines are highest.   We see the photoelectric absorption edges of both \ion{O}{viii} and \ion{O}{vii}, particularly during the periods of rapid variation when the contiuum intensity is high.  We detect the narrow radiative recombination continua of \ion{O}{viii} and \ion{O}{vii} at their respective absorption edges.  This means that the line excitation mechanism in \object{EXO~0748$-$67} is via radiative cascades following photoionization.    

While the intensity of the continuum level varies significantly, the shape of the continuum is constant.  In addition, the equivalent widths of the spectral lines vary with the continuum intensity, but the actual emission line fluxes remain constant thoughout the observations.  The absorption structure is most prominent when the continuum intensity is high, but the optical depths at the edges do not change.  We must therefore always be looking through the same material. 

\subsection{Velocity Broadening}

\begin{figure}
      \resizebox{8.5cm}{!}{\rotatebox{0}{\includegraphics{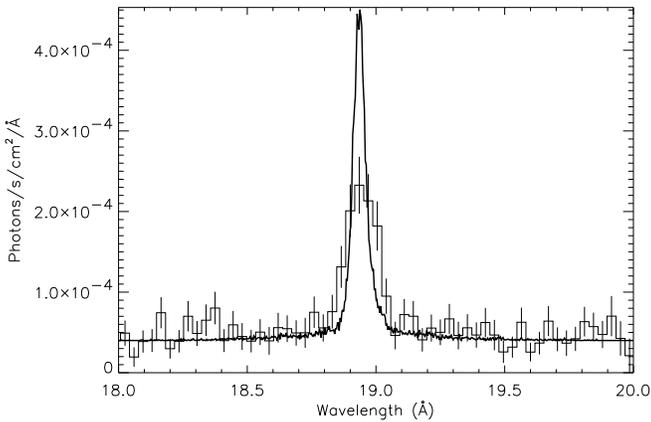}}}
      \caption[]{\ion{O}{viii} Ly$\alpha$.  The histogrammed data is overlaid with the normalized raytraced LSF.  The error bars on the data contain both the statistical errors and the $10\%$ uncertainty in the effective area. 
              }
         \label{FigVelocity}
   \end{figure}

All of the emission lines show significant velocity broadening.  This is illustrated in Fig. ~\ref{FigVelocity}, which shows the \ion{O}{viii} Ly$\alpha$ line overlaid with the instrument line spread function (LSF).  Fitting the Ly$\alpha$ lines by convolving the LSF with gaussian distributions, we measure velocity widths of $\sigma = (2600 \pm 490)\, {\rm km\, s^{-1}}$ for \ion{Ne}{x} Ly$\alpha$, $(1390 \pm 80)\, {\rm km\, s^{-1}}$ for \ion{O}{viii} Ly$\alpha$, and $(850 \pm 180)\, {\rm km\, s^{-1}}$ for \ion{N}{vii} Ly$\alpha$.  We find a direct correlation between the magnitude of the velocity width and the degree of ionization. 
We measure the shifts in the line centroids to be less than $0.020\, {\rm \AA}$ for all of the lines.  This means that the systemic velocity of the line emitting plasma with respect to the line of sight is less than $300\, {\rm km\, s^{-1}}$.  

\subsection{Density Sensitivity}

\begin{figure}
      \resizebox{8.5cm}{!}{\rotatebox{0}{\includegraphics{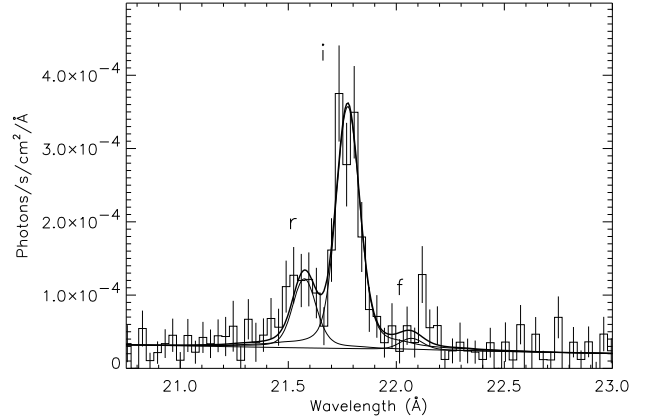}}}
      \caption[]{\ion{O}{vii} He-like series.  The instrument LSF, broadened to account for a $1390\, {\rm km\, s^{-1}}$ velocity field, is overlaid on the histogrammed data.  The contributions from the resonance line (r), intercombination lines (i), and forbidden line (f) are shown with thin lines.  The thick line shows the combined fit.  
              }
         \label{FigTriplet}
   \end{figure}

We can use the line ratios in the He-like complexes to put limits on the density of the plasma in the different ionization regions.  At sufficiently high densities the upper energy level of the forbidden line transition is collisionally depopulated through the upper level of the intercombination lines.  The ratio of the forbidden line flux to the intercombination line flux is therefore sensitive to the density of the plasma.  The He-like complexes in \object{EXO~0748$-$67} show the weak resonance lines characteristic of a photoionized plasma, bright intercombination lines and no obvious forbidden lines (see Fig. ~\ref{FigTriplet}).  We have measured upper-limits to the flux in the forbidden lines and compared them to the flux in the intercombination lines to determine lower-limits to the electron densities.  For \ion{O}{vii} we find an upper limit of $0.19$ for the ratio, which corresponds to a lower-limit to the electron density of $2.0\times 10^{12}\, {\rm cm^{-3}}$.  For \ion{Ne}{ix} we measure an upper-limit of $0.21$, which corresponds to a lower-limit of $7\times 10^{12}\,{\rm cm^{-3}}$ (Porquet \& Dubau \cite{Porquet}).  

\section{Discussion}

From the observed spectral behavior and the derived source parameters we can construct an empirical model of the circumstellar material.     
The large equivalent widths of the emission lines require that the emission regions subtend a fairly large solid angle in order to be sufficiently illuminated by the central source.  These regions must also extend high above the plane of the binary or else we would have observed eclipses or other modulations correlated with the orbital phase.  We can estimate the geometry of this material by comparing the observed absorption and emission in the hydrogen-like and helium-like oxygen ions.  For a plasma in photoionization equilibrium the ionization rate is equal to the recombination rate everywhere in the gas.  This means that the number of photons that are absorbed by a particular ion can be directly compared to the number of photons emitted by that ion.  Given the efficiency of cascade through a particular line transition, $\eta_{\rm line}$, this can be expressed as
\begin{equation} 
  N_{\rm line}=\eta_{\rm line} N_{\rm recombined} = \eta_{\rm line} N_{\rm absorbed} \,,
\end{equation} 
where $N$ is the photon flux.  The ratio of observed absorption to observed emission is then determined by the geometry of the system.  For example, for a disk oriented with its axis along the line of sight, photons will be absorbed in the surrounding material but then re-emitted isotropically, so that the observed line emission would be higher than expected from the observed absorption.   For \object{EXO~0748$-$67} we have compared the number of photons absorbed at the \ion{O}{viii} and \ion{O}{vii} edges to the number of photons emitted in the \ion{O}{viii} Ly$\alpha$ and \ion{O}{vii} He-like complexes.  We used cascade efficiencies ($44\%$ for \ion{O}{viii} and $75\%$ for \ion{O}{vii}) calculated with the atomic model for a recombining plasma used in Sako et al. (\cite{Sako}).  We see approximately three times more absorption than expected in each of the ions.  The absorbing material must therefore be flattened along our line of sight with a solid angle of roughly $\frac{1}{3} 4\pi$.  Since we know that the system is highly inclined, this material must be aligned with the plane of the binary and therefore with the accretion disk.  

The fact that the emission lines show large velocity broadening with no systemic Doppler-shifts suggests that the observed widths are not associated with net outflow or inflow but rather with orbital motion around the central source.  Using the ionization parameters of formation calculated as described in Sako et al. (\cite{Sako}) (see Table ~\ref{TabValues} for these values) we find a correlation with velocity that is consistent with an orbital velocity structure; \ion{Ne}{x}, which has the largest ionization parameter and is preferentially emitted closest to the central source, is observed to be moving with the largest velocity while \ion{N}{vii}, which has the lowest ionization parameter and should be emitted farthest from the source, is observed to be moving with the lowest velocity.     

The \object{EXO~0748$-$67} system must therefore contain a thickened accretion disk where the orbiting material extends high above the binary plane.  The plasma in the uppermost regions probably produces both the emission and absorption features in the observed spectra.  The continuum intensity is independent of orbital phase so the central source region must be extended; a compact X-ray source would show orbital modulation in such a highly inclined system.  Since the flux in the emission lines remains constant, the source luminosity must be constant; variations in luminosity would cause changes in the ionization structure and therefore changes in both the line flux and the line ratios.  The optical depths at the absorption edges are also constant so we must always be looking through the same absorbing and emitting material.  The observed variations in continuum intensity are best explained by local obscurations of the central source region.  
 Interestingly, the velocity parameters derived for the optically-emitting material ( $\sim 2000 \, {\rm km\, s^{-1}}$ line broadening with $210 \pm 92\, {\rm km\, s^{-1}}$ amplitude Doppler-shifts (Crampton et al. \cite{Crampton})) are very similar to those measured here.  Although it must be in a very different region of ionization, it is possible that the optically-emitting material is also orbiting in this thickened disk.

As a simple test of this model we can compare the emission measure for an orbiting plasma to the empirical emission measure.  Assuming a velocity profile with $v=(\frac{GM}{r})^{1/2}$ and a mass of $1.4 M_{\sun}$ for the neutron star, we can calculate the radial distances for each ion from the measured velocities.  Then, using the ionization parameters of formation ($\xi=\frac{L}{n_{\rm e} r^2}$) and the radial distances we can calculate the densities in these regions assuming a source luminosity of $1 \times 10^{36} {\rm ergs \, s^{-1}}$.  Defining the emission measure as 
\begin{equation} 
{\rm EM}=\int dVn_{\rm e}^2 \simeq \Delta \Omega r^{2} \Delta r n_{\rm e}^2 \,,
\end{equation} 
we take $\Delta r = \frac{1}{2} r$, and $\Delta \Omega = \frac{1}{2} \frac{1}{3} 4\pi$ to account for the estimated geometry.  The resulting emission measures as well as the components of the calculation are given in Table ~\ref{TabValues}.  We find emission measures that range from $\sim 22 \times 10^{57} {\rm cm}^{-3}$ for \ion{Ne}{x} Ly$\alpha$ to $\sim 10 \times 10^{57} {\rm cm}^{-3}$ for \ion{N}{vii} Ly$\alpha$.  

To calculate the empirical emission measure we divided the line luminosity by the line power evaluated at the ionization parameter of formation.  The line power for each ion was calculated using the HULLAC atomic codes (Bar-Shalom et al. \cite{BarShalom}) following the procedure described in Sako et al. (\cite{Sako}).   
We adjusted our measured line fluxes to account for interstellar absorption using the cross sections of Morrison \& McCammon (\cite{Morisson}) with a column density of $1.1 \times 10^{21} {\rm cm^{-3}}$.  We assumed a distance to the source of $D=10 \, {\rm kpc}$ (Gottwald et al. \cite{Gottwald}).  The resulting emission measure for each ion, assuming solar abundances (Anders $\&$ Grevesse \cite{Anders}), is included in Table ~\ref{TabValues}.  For \ion{Ne}{x} and \ion{N}{vii} we find emission measures of $2.1$ and $2.6 \times 10^{57} {\rm cm^{-3}}$.  For \ion{O}{viii} we find an emission measure of $0.6\times 10^{57} {\rm cm^{-3}}$.  This suggests an underabundance of oxygen that may have interesting implications for the evolutionary state of the companion star.  Adopting the factor of two uncertainty in the empirical emission measure that is estimated by Sako et al. (\cite{Sako}), we find that the predicted emission measures are roughly a factor of five higher than the empirical emission measure except at \ion{O}{viii}.  Considering that the predicted emission measures are highly dependent on our geometric assumptions and on a velocity profile that is only strictly valid for a thin Keplerian disk, the agreement with the empirical emission measures is remarkably good.  

This empirical model of a thickened disk is difficult to understand theoretically.  For a disk in hydrostatic equilibrium, the vertical height at a particular radius should be roughly given by the product of the radius and the ratio of the sound speed of the gas to its local orbital velocity.  Here, we can directly infer the sound speed from the ionization parameter and the orbital velocity from the widths of the lines, yet we infer a disk height that grossly violates this condition.  However, despite the theoretical challenge that this presents, we find no alternative model that satisfies the empirical constraints.  The large observed line equivalent widths and the clear evidence for intrinsic absorption essentially guarantee that this material is well out of hydrostatic equilibrium.  Hopefully, as high-resolution spectroscopic observations of other LMXB's accumulate, we may gain a clearer understanding of how such accretion flows can be formed and maintained.

\begin{acknowledgements}
We thank the rest of the RGS team, and in particular Frits Paerels and Masao Sako, for helpful conversations during this analysis.  This work is based on observations obtained with the {\it XMM-Newton}, an ESA science mission with instruments and contributions directly funded by ESA member states and the USA (NASA).  The Columbia University team is supported by NASA.  The SRON team is supported by the Netherlands Organization for Scientific Research (NWO).  JC acknowledges the support of a NASA Graduate Student Researcher's Program Fellowship.  
\end{acknowledgements}


\begin{table}
\begin{center}
\caption{Measured and Derived Parameters for the Ly$\alpha$ emission lines \label{tbl-2}}
\begin{tabular}{ccccccc}
\hline 
\hline 
Line & Velocity & Radius & ${\rm log}\xi$ & Density & ${\rm EM_{predicted}}$ & ${\rm EM_{empirical}}$ \\
     & (${\rm km\, s^{-1}}$) & (${\rm cm}$) & & (${\rm cm^{-3}}$) & ($10^{57}{\rm cm^{-3}}$) & ($10^{57}{\rm cm^{-3}}$) \\
\hline
Ne X   & $2600$ & $2.8\times 10^{9}$ & 2.1 & $2.0\times 10^{15}$ & $21.8$ & $2.1$ \\
O VIII & $1390$ & $9.6\times 10^{9}$ & 1.9 & $2.6\times 10^{14}$ & $15.8$ & $0.6$ \\
N VII  & $850$  & $2.6\times 10^{10}$ & 1.8 & $4.8\times 10^{13}$ & $10.3$ & $2.6$ \\
\hline
\label{TabValues}
\end{tabular}


\end{center}
\end{table}

\end{document}